\documentclass{article}
\usepackage{spconf,amsmath,graphicx,multirow}
\usepackage{todonotes}
\usepackage{url}
\usepackage{tabularx}
\usepackage{soul}
\usepackage{xcolor}
\usepackage{enumitem}

\title{Conformer-based Target-Speaker Automatic Speech Recognition for Single-Channel Audio}
\name{\begin{tabular}{c} Yang Zhang$^*$, Krishna C. Puvvada$^*$, Vitaly Lavrukhin, Boris Ginsburg
\thanks{*Equal contribution.}
\end{tabular}
}
\address{
  NVIDIA, USA}

\begin{document}
%\ninept
%
\maketitle

\begin{sloppypar}
\begin{abstract}
%  The abstract should contain about 100 to 150
% words

We propose CONF-TSASR, a non-autoregressive end-to-end time-frequency domain architecture for single-channel target-speaker automatic speech recognition (TS-ASR). The model consists of a TitaNet based speaker embedding module, a Conformer based  masking as well as ASR modules. These modules are jointly optimized to transcribe a target-speaker, while ignoring speech from other speakers. For training we use Connectionist Temporal Classification (CTC) loss and introduce a scale-invariant spectrogram reconstruction loss to encourage the model better separate the target-speaker's spectrogram from mixture. We obtain state-of-the-art target-speaker word error rate (TS-WER) on WSJ0-2mix-extr (4.2\%). Further, we report for the first time TS-WER on WSJ0-3mix-extr (12.4\%), LibriSpeech2Mix (4.2\%) and LibriSpeech3Mix (7.6\%) datasets, establishing new benchmarks for TS-ASR. The proposed model will be open-sourced through NVIDIA NeMo toolkit.

\end{abstract}

\begin{keywords}
Target-speaker ASR, Conformer, multi-speaker ASR, source separation
% , Multi-Speaker ASR
\end{keywords}

\section{Introduction}
\label{sec:intro}

\begin{figure}[t]
\includegraphics[width=\linewidth]{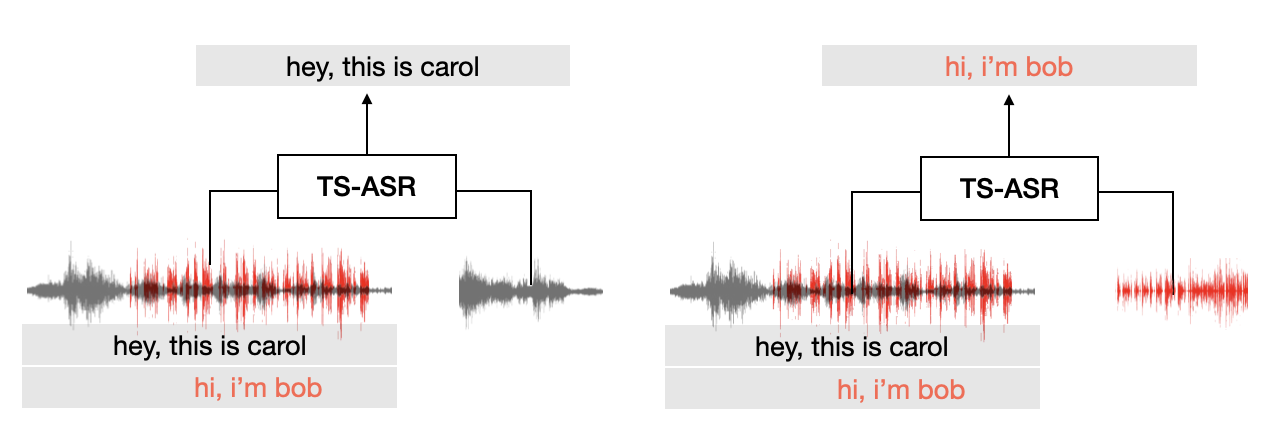}
\caption{Target-speaker ASR transcribes a specific speaker's part in a mixed utterance based on their clean speech sample (auxiliary utterance).}
\label{fig:task}
\end{figure}

% Overlapped speech recognition is a challenging aspect of practical ASR occuring during situations like diarization, speech recognition in presence of interfering speakers. 
Target-speaker automatic speech recognition (TS-ASR) is the task to transcribe a specific speaker's speech in an overlapping multi-speaker environment given the speaker's profile - an auxiliary utterance (Fig.~\ref{fig:task}).
  Along with blind source separation (BSS) and multi-speaker ASR, TS-ASR constitutes a class of approaches for overlapped speech recognition.

BSS methods separate individual components from a speech mixture in time-domain~\cite{luo2019conv, subakan2021attention} which are passed on to a single-speaker ASR model for transcription as a second step. As the separation step of BSS is not optimized for ASR, this can be sub-optimal. Multi-speaker ASR approaches~\cite{chang2019end, sklyar2021streaming, kanda2020serialized, guo2021multi} and their speaker-attributed variants (SA-ASR)~\cite{kanda2021end,kanda2022streaming} generate transcripts as output and are optimized end-to-end for ASR. A characteristic of BSS models and their analogous multi-speaker ASR models is their multiple output branches, one per source. SA-ASR requires profiles of all speakers in a mixed utterance as auxiliary information.

These seemingly similar approaches for overlapping speech recognition come with their own set of pros and cons. While BSS and their analogous multi-speaker ASR approaches do not require any auxiliary information, their major limitations include predefined number of output streams, permutation invariant training (PIT)~\cite{kolbaek2017multitalker} and speaker-tracing~\cite{speakerbeam} for long audio inference. Further, having a different number of speakers during training and inference can greatly reduce their performance. In case of multi-speaker ASR, serialized output training (SOT) can overcome some of these limitations, but leaves much to be desired for in terms of performance~\cite{kanda2020serialized}. In conjunction with SOT, SA-ASR uses speaker profile information to improve performance and not be limited by fixed number of outputs. But, it assumes availability of profiles for \emph{all} speakers in an utterance. Nonetheless, it is well suited for transcribing meeting like scenarios. TS-ASR~\cite{speakerbeam,moriya2022streaming}, on the other hand, requires only one speaker profile of interest. It is apt for situations that require transcribing one target-speaker while ignoring interfering speakers. By design, TS-ASR doesn't suffer from permutation ambiguity and speaker-tracing. However, it requires one inference per speaker if used to transcribe multiple speakers.

%%The task is general enough to be used for both speech diarization of meetings with partially overlapped speech at a similar signal to noise ratio and speech enhancement of a foreground speaker in a noisy environment with interfering speech in the background. 
% -------------------------------------------------------------------------

In this paper, we propose Conformer-based TS-ASR model (CONF-TSASR) to address single-channel target-speaker ASR. Our approach adopts the SpeakerBeam~\cite{speakerbeam} architecture and makes the following contributions:
\begin{itemize}
\itemsep-0.1cm
    \item Improve SpeakerBeam using TitaNet~\cite{koluguri2022titanet} and Conformer~\cite{gulati2020conformer} modules trained in an end-to-end fashion with CTC and novel spectrogram reconstruction loss.
    \item Achieve state-of-the-art target-speaker WER (TS-WER) on WSJ0-2mix-extr$^2$ and report results on WSJ0-3mix-extr and LibriSpeechMix~\cite{kanda2020joint} for the first time.
    \item Study the effects of target-speaker SNR and length of auxiliary utterance on model performance. 
\end{itemize}
% Model will be open-sourced through NVIDIA NeMo toolkit\footnote{\url{https://github.com/NVIDIA/NeMo}}.
% show by updating the original SpeakerBeam we can achieve state-of-the-art results on de-facto benchmarks, such as WSJ0-mix~\cite{wsj0} and LibriSpeechMix~\cite{kanda2020joint}. Apart from the ASR loss with added a spectrogram loss. We evaluated the model on multiple benchmarks. 
% Our model CONF-TSASR learns a Conformer-based~\cite{gulati2020conformer} masking network as well as a Conformer-CTC based ASR model conditioned on TitaNet~\cite{koluguri2022titanet} -- an embedding network based on Contextnet~\cite{han2020contextnet}.
% We compare our one-pass model with SpeakerBeam and related multi-speaker models and show that it is superior to a cascading model consisting of Exformer and a Conformer-CTC based ASR model. 

% related work like blind speech separation speakerbeam, exformer, and end-to-end transformer 

\section{Conformer-based TS-ASR Architecture}
\label{sec:tsasr}

\begin{figure}[t]
\centering
\includegraphics[width=0.8\linewidth]{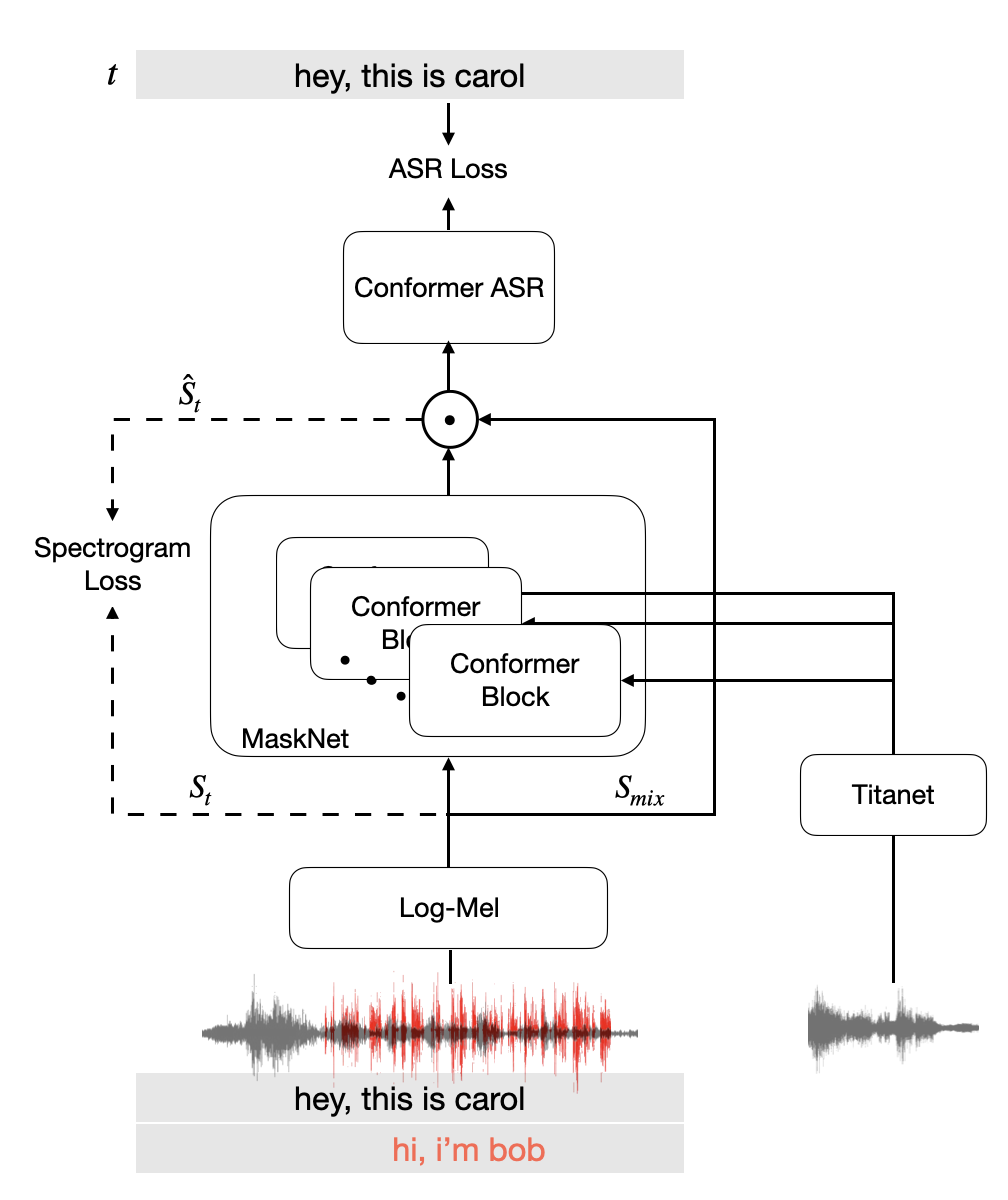}
\caption{Conformer-based CONF-TSASR model architecture. Feature extraction creates log-Mel spectrogram $S_{mix}$ of the mixed utterance. A speaker embedding is extracted from an auxiliary utterance using TitaNet. The masking network learns a time-frequency mask for the target-speaker. The model is trained using CTC-loss and spectrogram reconstruction loss using the target-speaker's individual spectrogram $S_t$.}
\label{fig:model}
\end{figure}

The proposed single-channel CONF-TSASR model consists of three modules - TitaNet, MaskNet and an ASR module (Fig.~\ref{fig:model}). It takes two inputs - a mixed utterance and a clean auxiliary utterance from the target-speaker and transcribes only the target-speaker's speech from the mixed utterance.  The auxiliary utterance is encoded into a 192-dim speaker embedding by TitaNet~\cite{koluguri2022titanet} – a speaker embedding extractor model based on ContextNet~\cite{han2020contextnet}. From the mixed utterance 80-dim log-Mel features are extracted every 10msec over a window of 25msec. These are further perturbed with SpecAugment~\cite{park2019specaugment} and sub-sampled by 4x using two convolutional layers. MaskNet takes the sub-sampled features ($S_{mix}$) and speaker embedding to produce a time-frequency mask which is multiplied with $S_{mix}$ to estimate the target-speaker's time-frequency features ($\hat{S_t}$). Finally, a Conformer~\cite{gulati2020conformer} ASR module is used to transcribe the target speaker’s speech using $\hat{S_t}$. The entire model is optimized using CTC~\cite{ctc} loss and spectrogram reconstruction loss. The latter computes scale invariant SiSNR~\cite{le2019sdr} between an up-sampled $\hat{S_t}$ -- the estimated spectrogram -- and true spectrogram $S_t$.  The spectrogram reconstruction loss is reserved for training where you have access to the individual sources. 
%For example, when simulating the mixed utterance using dynamic mixing mixing~\cite{zeghidour2021wavesplit}.  }

In our experiments, both MaskNet and ASR module consist of 18 Conformer layers, each with a hidden dimension of 256 and feed-forward dimension of 1024. Multi-head attention consists of 4 heads and the kernel size of the convolutional module is 31. The speaker embedding is linearly projected to match MaskNet's hidden dimension of 256 and added to the input of every Conformer block.
Both ASR module and TitaNet are initialized with pre-trained weights available in NVIDIA NeMo toolkit\footnote{\url{https://github.com/NVIDIA/NeMo}}. For TitaNet, we freeze its ContextNet encoder and only train the decoder.
The CONF-TSASR model has 66.1M trainable parameters and 85.4M parameters in total including the frozen TitaNet encoder.

\section{Experiments}
\label{sec:experiments}

\subsection{Datasets}
\label{sec:datasets}

We evaluated the proposed approach using two and three speaker mixtures created from WJS0~\cite{wsj0} and Librispeech~\cite{librispeech} datasets following \cite{hershey2016deep} and \cite{kanda2020joint} respectively. To adapt these mixture datasets for TS-ASR, we augment them with a random auxiliary utterance from the target-speaker\footnote{\url{https://github.com/xuchenglin28/speaker_extraction}}. 
Briefly, a training example for two-speaker mixture was created by first randomly selecting two speakers and choosing one as the target-speaker. Among all utterances by the target-speaker two are chosen, one for creating the auxiliary utterance and one for the mixed utterance. To create the mixture we pick an utterance spoken by the other speaker. For WSJ0 mixtures, utterances were combined at a SNR uniformly chosen between 0 and 5 dB for each mixture~\cite{hershey2016deep}. For Librispeech mixtures, chosen utterances were combined without changing SNR to be consistent with \cite{kanda2020joint}. For WSJ0 mixtures, the shorter utterance is both prepended and appended with random length of silence to match the length of the longest utterance in the mixture. In contrast, Librispeech mixtures are generated following \cite{kanda2020joint}, where the utterances were shifted by random delays before being added to the mixture. Delay values were chosen under the constraint that the start times of each utterance differed by 0.5sec or longer. Further, we augment the training data using speed and volume perturbation. For speed perturbation~\cite{speechaugment_interspeech}, the speed of each individual utterance is modified with a probability of 0.3 from its original rate to 95\%, 97.5\%, 100\%, 102.5\% or 105\% before mixing. Volume perturbation \cite{kanda2020joint} involves scaling the final mixture’s volume by a random factor sampled from $[0.125, 2.0]$. In the following, we refer to two and three speaker mixtures of WSJ0 as WSJ0-2mix-extr and WSJ0-3mix-extr respectively, whereas LibriSpeech2mix and LibriSpeech3mix denotes Librispeech mixtures.

% \begin{figure}[t]

% \begin{minipage}[b]{.48\linewidth}
%   \centering
%   \centerline{\includegraphics[width=4.0cm]{images/overlap_wsj2.png}}
% %  \vspace{2.0cm}
%   \centerline{(a) WSJ0-2mix}\medskip
% \end{minipage}
% \hfill
% \begin{minipage}[b]{.48\linewidth}
%   \centering
%   \centerline{\includegraphics[width=4.0cm]{images/overlap_wsj3.png}}
% %  \vspace{1.5cm}
%   \centerline{(b) WSJ0-3mix}\medskip
% \end{minipage}
% %
% \begin{minipage}[b]{.48\linewidth}
%   \centering
%   \centerline{\includegraphics[width=4.0cm]{images/overlap_ls2.png}}
% %  \vspace{1.5cm}
%   \centerline{(c) LibriSpeech2Mix}\medskip
% \end{minipage}
% \hfill
% \begin{minipage}[b]{0.48\linewidth}
%   \centering
%   \centerline{\includegraphics[width=4.0cm]{images/overlap_ls3.png}}
% %  \vspace{1.5cm}
%   \centerline{(d) LibriSpeech3Mix}\medskip
% \end{minipage}
% %
% \caption{Speaker overlap distribution of WSJ0-mix (\textit{max} version) and LibriSpeechMix.}
% \label{fig:overlap}
% \end{figure}

\subsection{Training Setup}
\label{sec:training_setup}

For training, we used 16 V100-32GB GPUs with a global batch size of 64. The model was trained for 60K and 480K updates for WSJ0-mix-extr and LibriSpeechMix respectively. We used AdamW with a peak learning rate of $3*10^{-4}$ and 0.01 weight decay. We used 10K and 25K warmup steps respectively with Cosine annealing and a minimum learning rate of $1*10^{-6}$. The relative weights of losses, when more than one is used, are tuned based on validation TS-WER.

\begin{figure}[t]
\centering
\includegraphics[width=0.75\linewidth]{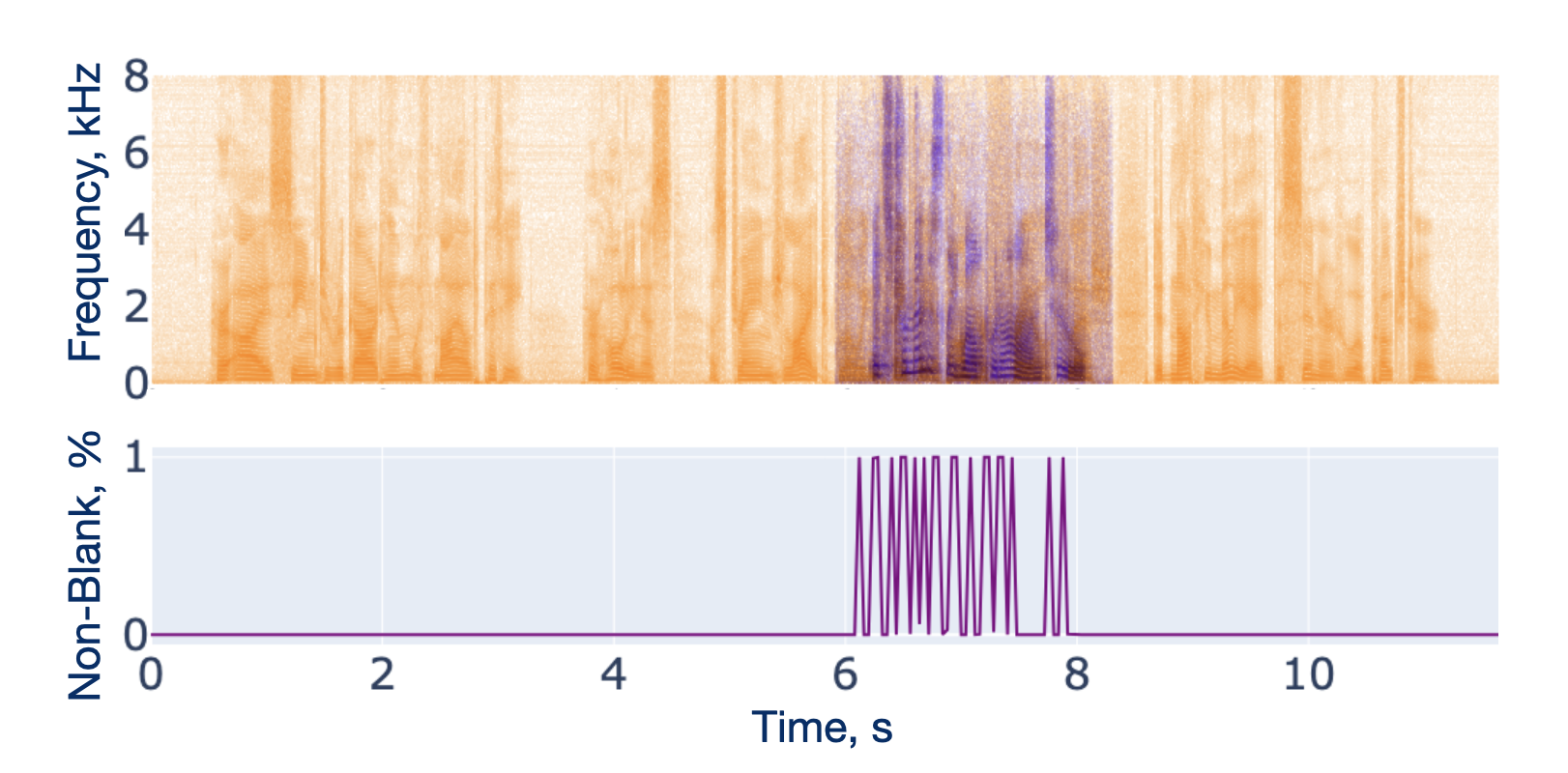}
\caption{A LibriSpeech2mix example. (top) Input mixture spectrogram. Target-speaker and overlapping speaker are shown in different colors. (bottom) Time-aligned non-blank emission probabilities for target-speaker using the proposed model. (Best viewed in color.)}
\label{fig:mixture_librispeech}
\end{figure}

\subsection{WSJ0-mix-extr results}
\label{sec:results_wsj}

% actual params were 85.4 - rounded off to 85
\begin{table*}[t]
    \centering
    \begin{tabular}{c|c|ccccc}
     Model       & Params, M & N &  Loss & Learn Embedding   & 2-mix & 3-mix\\
     % & & &Embedding & &\\
    \hline
                 &          & 2 & CTC & no & 8.6 & - \\
                 &          & 2 & CTC & yes & 4.8 & - \\
     CONF-TSASR  & 85       & 2 & CTC+Spec & yes  &   \textbf{4.2} & - \\
                 &          & 3 & CTC & yes & 5.4 &  13.8 \\
                 &          & 3 & CTC+Spec & yes  &  \textbf{4.8} & \textbf{12.4} \\
    \hline
    SpeakerBeam~\cite{speakerbeam} & n/a  & 2 &  Cross Entropy & yes &  30.6 & - \\
    Exformer~\cite{wang2022semi} + Conformer-CTC & 80 & 2 & SiSNR, CTC & no &  13.2 & - \\
    Conditional-Conformer-CTC~\cite{guo2021multi}  & n/a & 3 & CTC & yes &  19.9$^{**}$ & 34.3$^{**}$ \\
    Conformer-CTC  & 29 & 1 & CTC & no &  36.7$^*$ & 54$^*$ \\
    \hline
    \end{tabular}
    \caption{TS-WER of different models on WSJ0-2mix-extr and WSJ0-3mix-extr. The model was trained on N-maximum speakers in train mixture. $^*$ denotes best WER on individual transcript.  $^{**}$ denotes WER for multi-speaker ASR model.}
    \label{tab:wsj}
\end{table*}

\begin{table}[t]
\centering
\begin{tabular}{ c|c|cc }
 Model    & Params, M & L (sec)  & 2-mix\\
     \hline
 CONF-TSASR CTC  & 85 & \begin{tabular}{@{}c@{}}7.5 \\ 15\end{tabular}  &  \begin{tabular}{@{}c@{}}5.1 \\ 4.6\end{tabular}\\
 CONF-TSASR CTC+Spec  & 85 & \begin{tabular}{@{}c@{}}7.5 \\ 15\end{tabular}  &   \begin{tabular}{@{}c@{}}4.5 \\ \textbf{4.2} \end{tabular}\\
     \hline
SOT-Confomer-AED~\cite{kanda2021end}   & 129 & 15 &   4.9$^\dagger$ \\
t-SOT TT-18~\cite{kanda2022streaming}& 82 & 15  &  5.2$^\ddagger$  \\ 
t-SOT TT-36~\cite{kanda2022streaming}& 139 & 15  &  4.4$^\ddagger$  \\
\hline
\end{tabular}
\caption{TS-WER of CONF-TSASR, SA-WER$^\dagger$, and permutation invariant SA-WER$^\ddagger$ of related multi-speaker references on LibriSpeechMix, trained for up to 2 speakers. Notation: L - Length of auxiliary utterance}
\label{tab:librispeech2mix}
\end{table}

% \begin{table}[t]
% \centering
% \begin{tabular}{ c|c ccc }
%  Model   & L (sec) & 2-mix & 3-mix\\
%      \hline
%  CONF-TSASR CTC  & \begin{tabular}{@{}c@{}}7.5 \\ 15\end{tabular} & \begin{tabular}{@{}c@{}}7 \\6 \end{tabular} & \begin{tabular}{@{}c@{}}9.7 \\8.4 \end{tabular} \\  CONF-TSASR CTC+Spec & \begin{tabular}{@{}c@{}}7.5 \\ 15\end{tabular} & \begin{tabular}{@{}c@{}}6.3 \\\textbf{5.4} \end{tabular} & \begin{tabular}{@{}c@{}}9 \\\textbf{7.6} \end{tabular} \\ 
%      \hline
% SOT-Confomer-AED~\cite{kanda2021end} &15 &  6.8$^\dagger$ & 9.6$^\dagger$ \\
% SOT-Confomer-AED~\cite{kanda2021end} SD &  15 &   \textbf{6.4}$^\dagger$ &  \textbf{8.5}$^\dagger$ \\
% \hline
% \end{tabular}
% \caption{TS-WER of CONF-TSASR, and permutation invariant SA-WER$^\ddagger$ of related multi-speaker references on LibriSpeechMix. CONF-TSASR was trained for up to 3 speakers. Notation: L - length of auxiliary utterance, SD - speaker deduplication~\cite{kanda2021end}}
% \label{tab:librispeech3mix}
% \end{table}

 \begin{figure}[t]
\centering
\includegraphics[width=0.62\linewidth]{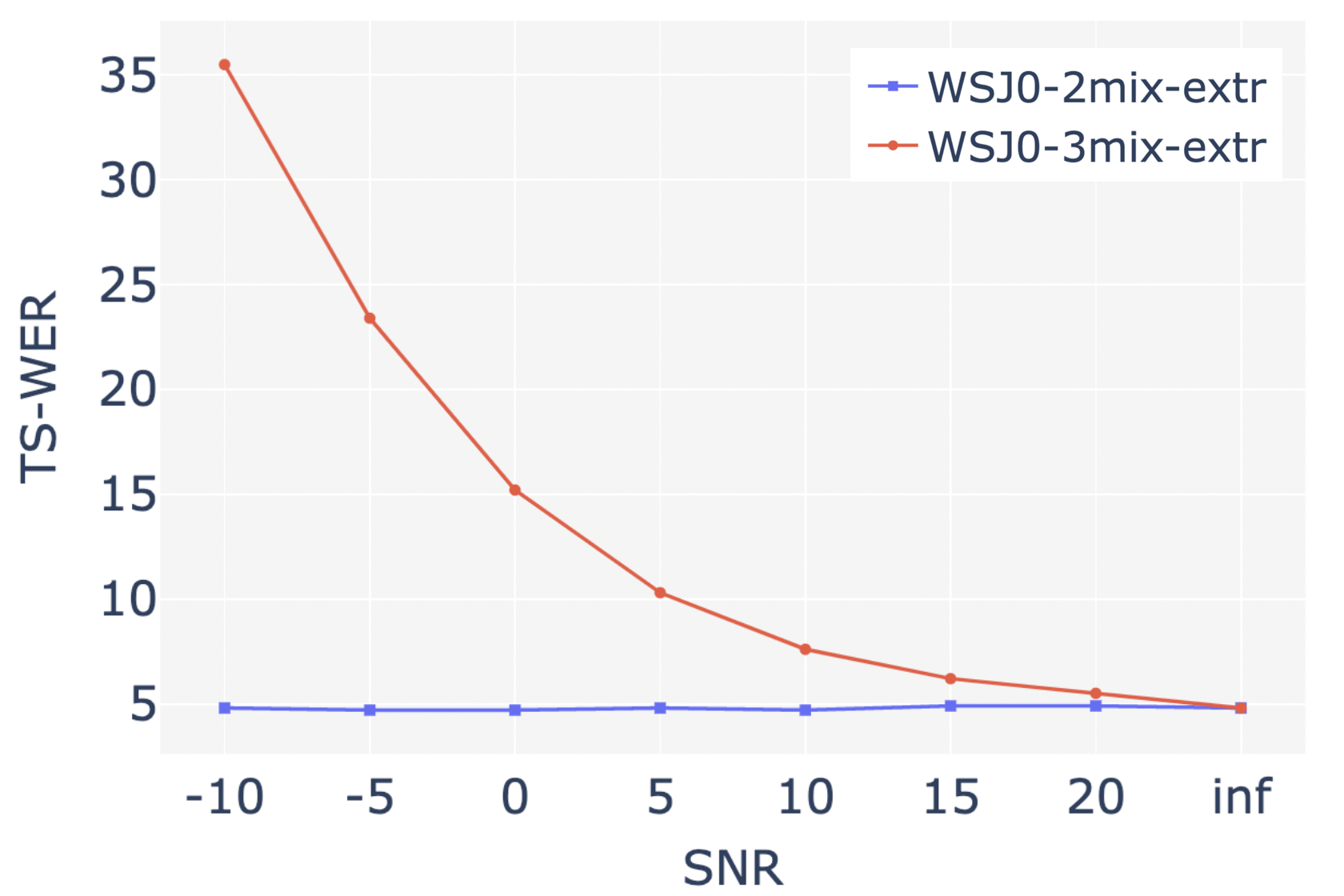}
\caption{TS-WER on WSJ0-2mix-extr and WSJ0-3mix-extr under different SNRs between target-speaker and interfering speakers using CONF-TSASR.}
\label{fig:snr_wsj}
\end{figure}

Table~\ref{tab:wsj} compares performance of the proposed CONF-TSASR model with contemporary results on the WSJ0-mix-extr datasets. For baselines, we include a conventional ASR model (Conformer-CTC),  SpeakerBeam~\cite{speakerbeam}, Exformer~\cite{wang2022semi} and Conditional-Conformer-CTC~\cite{guo2021multi}. The first model was trained on single-speaker, the second and third were trained on two speakers and the last model was trained on up to three speakers. SpeakerBeam can be regarded as a TS-ASR model which directly generates transcription for target-speaker. Exformer is a state-of-the-art target-speaker source separation model
% \textcolor{red}{describe what is source extraction model either here or before} 
based on SepFormer~\cite{subakan2021attention} and thus requires an additional step to transcribe the model output. We updated the Exformer architecture with recent advancements to facilitate fairer comparisons. Namely, we replaced the original pre-trained embedding network in Exformer with the same pre-trained TitaNet~\cite{koluguri2022titanet} that was used in the CONF-TSASR. We also matched model size of the SepFormer with masking network in CONF-TSASR. Note that the model used to transcribe Exformer output and the "conventional ASR" baseline are the same. This in-turn is same as the one used for initializing ASR block in CONF-TSASR, except the former is further fine-tuned on training partition of WSJ0 dataset. Conditional-Conformer-CTC is a multi-speaker ASR model that uses conditional speaker chain to transcribe every speaker subsequently.

As expected, the conventional ASR model trained on single-speaker data performs poorly on WSJ0-mix (36.7\% WER on WSJ0-2mix), whereas SpeakerBeam and Exformer+Conformer-CTC show 30.6\% and 13.2\% TS-WER respectively. Conditional-Conformer-CTC reaches a WER of 19.9\% on WSJ0-2mix and 34.3\% on WSJ0-3mix. In comparison, CONF-TSASR trained on up to two-speaker mixtures reaches 4.8\% TS-WER and 4.2\% TS-WER with additional spectrogram reconstruction loss.   Exformer results show optimizing for SiSNR does not necessarily give the best result for transcription. Also, shifting away from time-frequency domain to time-domain is not only unnecessary for target-speaker speech recognition but also decreases model efficiency due to more time steps.
CONF-TSAR model reaches TS-WER of 4.8\% on WSJ0-2mix-extr and 12.4\% on WSJ0-3mix-extr when trained on up to three speakers, suggesting that the proposed model is able to transcribe target-speaker from single-channel input in spite of two distracting speakers.  To our knowledge, this is the best TS-WER reported on WSJ0-2mix-extr and the first study to report TS-WER on  WSJ0-3mix-extr.

Fig.~\ref{fig:snr_wsj} shows the sensitivity of CONF-TSASR w.r.t. SNR between target and overlapping speakers. We observe that the performance is more sensitive to SNR when the mixture contains three overlapping speakers compared to just two speakers.

\subsection{LibriSpeechMix results}
\label{sec:results_librispeechmix}
Tables~\ref{tab:librispeech2mix} \& \ref{tab:librispeech3mix} report TS-WER results for the first time on LibriSpeechMix datasets. 
%Due to lack of direct comparison in literature, we use
Due to lack of previous TS-ASR results on LibriSpeechMix dataset in literature, we use
SOT-Conformer-AED~\cite{kanda2021end} and its streaming version t-SOT~\cite{kanda2022streaming} as reference.\footnote{While \cite{moriya2022streaming} tackles TS-ASR, our work is not directly comparable to theirs as they focuses on streaming and evaluate on Japanese corpus.} These are different, yet closely related multi-speaker transcription models. They are based on transformer encoder-decoder architecture and use SOT~\cite{kanda2020serialized}. They differ with the proposed model in the following (non-exhaustive) ways.  1) They transcribe all speakers in a given utterance. 2) Have knowledge of speaker profiles for all possible speakers in a given utterance. 3) Use 10 auxiliary utterances during training and 2 during evaluation (each with avg. length of 7.5 sec) to calculate speaker profiles. 4) SOT-Conformer-AED reports speaker-attributed WER (SA-WER)~\cite{kanda2020joint} on LibriSpeech2Mix and LibriSpeech3Mix, while t-SOT reports permutation-invariant SA-WER~\cite{kanda2022streaming} on LibriSpeech2Mix. In contrast, the proposed model 1) Transcribes only target-speaker, 2) Is not aware of profiles for other speakers in utterance, 3) Uses only one auxiliary utterance during training and two during evaluation to calculate speaker profiles, and 4) Reports WER only for target speaker (TS-WER).

% \st{We believe that TS-ASR, having not seen all possible speaker profiles in an utterance, is more challenging task than multi-speaker ASR where the model knows that every token in utterance should be assigned to one of the speakers known apriori. (i think this last statement can be too controversial and should probably be removed. A counter argument could be: one mistake in multi-task ASR is counted as two in WER calculation - an insertion for one speaker and deletion from another speaker, but not in ours. We don't want to anger any reviewer}.
 To make the TS-WER results on LibriSpeechMix comparable to SA-WER, we transcribe all speakers in a mixed utterance with each speaker as target, one at a time, with CONF-TSASR model.
CONF-TSASR trained on up to three speakers achieves 5.4\% TS-WER on LibriSpeech2Mix and 7.6\% on LibriSpeech3Mix (Table~\ref{tab:librispeech3mix}). When trained on only up to two-speaker mixtures, the performance improves to 4.2\% TS-WER on LibriSpeech2Mix (Table~\ref{tab:librispeech2mix}). Both two and three-speaker results show that adding spectrogram loss (CTC+Spec) significantly outperforms using merely CTC loss. When evaluated using only one auxiliary utterance (7.5 sec) as speaker profile, the proposed model exhibits slight performance deterioration (e.g. 7.6\% vs. 9\%  in Table~\ref{tab:librispeech3mix}) highlighting the importance of robust speaker profiles. Training model with CTC loss~\cite{ctc} provides an auxiliary benefit of obtaining time-aligned token output probabilities for target-speaker (Fig.~\ref{fig:mixture_librispeech}, bottom).

\begin{table}[t]
\centering
\begin{tabular}{ c|c ccc }
 Model   & L (sec) & 2-mix & 3-mix\\
     \hline
 CONF-TSASR CTC  & \begin{tabular}{@{}c@{}}7.5 \\ 15\end{tabular} & \begin{tabular}{@{}c@{}}7 \\6 \end{tabular} & \begin{tabular}{@{}c@{}}9.7 \\8.4 \end{tabular} \\  CONF-TSASR CTC+Spec & \begin{tabular}{@{}c@{}}7.5 \\ 15\end{tabular} & \begin{tabular}{@{}c@{}}6.3 \\\textbf{5.4} \end{tabular} & \begin{tabular}{@{}c@{}}9 \\\textbf{7.6} \end{tabular} \\ 
     \hline
SOT-Confomer-AED~\cite{kanda2021end} &15 &  6.8$^\dagger$ & 9.6$^\dagger$ \\
SOT-Confomer-AED~\cite{kanda2021end} SD &  15 &   \textbf{6.4}$^\dagger$ &  \textbf{8.5}$^\dagger$ \\
\hline
\end{tabular}
\caption{TS-WER of CONF-TSASR, and permutation invariant SA-WER$^\ddagger$ of related multi-speaker references on LibriSpeechMix. CONF-TSASR was trained for up to 3 speakers. Notation: L - length of auxiliary utterance, SD - speaker deduplication~\cite{kanda2021end}}
\label{tab:librispeech3mix}
\end{table}

\section{Conclusion}
\label{sec:conclusion}

We present CONF-TSASR, an end-to-end state-of-the-art single-channel target-speaker speech recognition model. % based on Conformer architecture.
It consists of three modules. 
% speaker embedding, MaskNet, and ASR backbone. 
TitaNet -- extracts a speaker embedding from a target-speaker's auxiliary utterance.  MaskNet -- generates a time-frequency mask for a target-speaker using Conformer. ASR module -- transcribes the masked speech features using Conformer. The model is trained with CTC and spectrogram loss. 
We obtain state-of-the-art results on WSJ0-2mix-extr and establish new benchmarks for WSJ0-3mix-extr and LibriSpeechMix datasets.
The model can be both used for fully and partially overlapped speech, requires as little as one auxiliary utterance and is non-autoregressive.
Model will be open-sourced through NVIDIA NeMo toolkit\footnote{\url{https://github.com/NVIDIA/NeMo}}.

\vfill\pagebreak

\bibliographystyle{IEEEbib}
\bibliography{main}

\end{sloppypar}
\end{document}